\documentclass[manuscript,screen,nonacm]{acmart}
\usepackage{booktabs}
\usepackage{graphicx}
\usepackage{tabularx}
\usepackage{longtable}
\makeatletter
\@twosidefalse
\@mparswitchfalse
\renewcommand\@titlefont{\fontsize{14}{16}\selectfont\bfseries}
\makeatother
\renewcommand\footnotetextcopyrightpermission[1]{}
\graphicspath{{figures/}}
\newcommand{\nrows}{250}
\newcommand{\npos}{78}
\newcommand{\nneg}{172}
\newcommand{\ncovered}{39}
\newcommand{\nagents}{40}
\newcommand{\ntotalaa}{50}

\newcommand{\npages}{24}
\newcommand{\nplatforms}{11}

\newcommand{\ndevrows}{27}
\newcommand{\ndevpages}{19}
\newcommand{\nheldrows}{55}
\newcommand{\nheldpos}{15}

\newcommand{\abstainn}{92}
\newcommand{\committedn}{158}
\newcommand{\axeprecision}{0.90}
\newcommand{\axerecall}{0.36}

\newcommand{\vlmpureprecision}{0.61}
\newcommand{\vlmpurerecall}{0.67}
\newcommand{\vlmpurefone}{0.64}

\newcommand{\vlmcuedfone}{0.69}
\newcommand{\agentprecision}{0.56}
\newcommand{\agentrecall}{0.86}
\newcommand{\agentfone}{0.68}

\newcommand{\pcrecall}{0.59}

\newcommand{\claudeNaiveR}{0.60}

\newcommand{\claudeParityR}{0.77}

\newcommand{\codexNaiveR}{0.32}

\newcommand{\codexParityR}{0.54}

\newcommand{\unionP}{0.54}
\newcommand{\unionR}{0.95}

\newcommand{\unionTP}{74}
\newcommand{\unionFP}{63}
\newcommand{\intersectionP}{0.66}
\newcommand{\intersectionR}{0.58}

\newcommand{\intersectionTP}{45}
\newcommand{\intersectionFP}{23}
\newcommand{\costPure}{6.38}
\newcommand{\costCued}{7.29}
\newcommand{\costPc}{14.98}

\newcommand{\costClaudeCodeParity}{74.52}
\newcommand{\costCodexNaive}{32.76}

\newcommand{\costGemini}{36.65}

\newcommand{\costGpt}{64.19}

\newcommand{\pcAbstainPct}{36.8}

\newcommand{\candidatePrecision}{0.78}
\newcommand{\candidateRecall}{0.90}
\newcommand{\candidatePos}{105}
\newcommand{\repeatSame}{241}
\newcommand{\repeatAgreement}{96.4}

\setcopyright{none}
\begin{document}
\title{Agentic Web Accessibility Auditing: A Criterion-Specific Framework for Translating WCAG Requirements into Assessments}
\author{Arjun Mishra}
\affiliation{%
  \institution{University of British Columbia}
  \city{Vancouver}
  \state{BC}
  \country{Canada}}
\email{arjun03@student.ubc.ca}

\author{Pranav Karthik}
\affiliation{%
  \institution{University of British Columbia}
  \city{Vancouver}
  \state{BC}
  \country{Canada}}
\email{pmruthyu@student.ubc.ca}

\author{Byungjun Bae}
\affiliation{%
  \institution{Electronics and Telecommunications Research Institute}
  \city{Daejeon}
  \country{Republic of Korea}}
\email{1080i@etri.re.kr}

\author{Dongwook Yoon}
\affiliation{%
  \institution{University of British Columbia}
  \city{Vancouver}
  \state{BC}
  \country{Canada}}
\email{yoon@cs.ubc.ca}

\makeatletter
\if@ACM@anonymous\else
\renewcommand{\shortauthors}{Mishra et al.}
\fi
\makeatother
\begin{abstract}
\emph{\textbf{ABSTRACT}}
Web accessibility auditing requires interpreting diverse requirements and examining interface behavior. Rule-based checks and noninteractive model assessments can miss barriers requiring contextual or interactive evidence. We present an agentic framework that assigns a vision-language agent to each accessibility requirement. Guided by tailored instructions, agents inspect webpages, operate controls, and record evidence supporting their findings. We implement the framework for 40 requirements from the Web Content Accessibility Guidelines (WCAG). To compare detection and cost, we construct a dataset of 250 page--criterion records derived from expert audits across 11 scholarly platforms. Agents recover 67 of 78 reported positive cases (86\% recall), compared with 36\% for axe-core, a rule-based checker, and 67\% for an uncued, noninteractive vision-language model, at lower precision (56\%). They recover nine of ten Keyboard and No Keyboard Trap cases missed by both baselines. Together, the framework, implementations, and dataset support automated accessibility auditing grounded in inspectable evidence.
\end{abstract}


\maketitle
\pagestyle{plain}
\section{Introduction}
Web accessibility evaluation asks whether people with different access requirements can perceive, understand, and operate an interface. A missing accessible name (text identifying an element to assistive technology), an unreadable contrast combination, and a component that cannot be left using the keyboard create different barriers and require different forms of investigation. The Web Content Accessibility Guidelines (WCAG) organize testable requirements into success criteria and conformance levels~\cite{w3cwai}. A success criterion specifies a requirement that content must satisfy; Level AA conformance includes the applicable Level A and AA requirements. These criteria provide a common vocabulary for documenting findings, but satisfying a checklist does not exhaust the accessibility problems people encounter. Research with blind web users demonstrates the importance of considering experienced barriers alongside guideline-based evaluation~\cite{power2012}. We study automation as support for this broader evaluation process.

Consider a search interface with a labelled input, a filter panel, and results. Initial inspection might establish that the input has an accessible name and the results have headings. It would not show whether opening the filter moves focus appropriately or whether a keyboard user can return to the results. Conversely, exercising the filter would not establish that its labels communicate their purpose. This example explains why assessment procedures must match the requirement: a page-level summary can obscure which questions were examined.

An automated auditor's conclusions depend on the evidence it can obtain. A rule-based engine can identify many violations with a reproducible decision procedure, such as an image element without the required naming information. Other decisions depend on meaning: the presence of alternative text does not establish that it conveys the purpose of an image. Still others concern behavior, including whether a control can be reached, activated, and left with a keyboard. These distinctions concern particular checks within criteria rather than disjoint classes of criteria. For example, source inspection may identify suspicious focus attributes, whereas observing a focus transition requires operating the rendered interface. Consequently, a tool can offer useful partial coverage of a criterion without establishing that the criterion is satisfied in every relevant state~\cite{vigo2013,gena11y2025}.

Recent language-model systems extend evaluation beyond fixed rules. GenA11y extracts criterion-relevant elements for model assessment; AccessGuru combines detection and correction using web-content representations~\cite{gena11y2025,fathallah2025accessguru}. Interactive testing offers a complementary approach. BAGEL investigates web keyboard barriers, while TaskAudit executes mobile tasks through a modified mobile screen reader (TalkBack) that executes gestures and captures output transcripts and analyzes the resulting evidence~\cite{bagel2023,zhong2026taskaudit}. These precedents motivate our question: how can a shared agent framework organize evidence collection and decisions across named WCAG criteria, and what does its evaluation reveal about its capabilities and limitations?

We address this question through a framework that assigns one \emph{worker agent} to each targeted success criterion. A worker is a model-driven process that receives a criterion-specific instruction, selects tools, interprets their observations, and returns a structured assessment. A \emph{tool} is an executable operation available to that process, such as querying elements, capturing a screenshot, measuring contrast, or sending keyboard input. Workers share a control loop and tool implementations, while their evidence requirements and decision instructions vary by criterion. This decomposition makes the relationship between a requirement, an action, and a reported finding explicit. It also provides a practical unit for inspecting failures: an unsuccessful assessment can be examined in the context of the particular requirement the worker was assigned.

We instantiate \nagents{} workers, including \ncovered{} of the \ntotalaa{} WCAG 2.1 Level A and AA criteria and one additional WCAG 2.2 criterion. Implementation breadth is one outcome of the work; demonstrated detection coverage is another. Our primary evaluation contains positive reference cases for 15 criteria, so it cannot establish violation-detection performance for every implemented worker. We investigate the framework using page--criterion records derived from Library Accessibility Alliance (LAA) audit reports~\cite{laa2024}. Reported findings provide positive reference labels, while the absence of a listed finding supplies an inferred negative. This construction enables comparisons of reported positives and negatives, but it does not turn a high-level audit report into an exhaustive assessment of every saved page state.

The study is organized around four research questions. \textbf{RQ1, Architecture:} how can shared control and tools be combined with criterion-specific instructions to produce inspectable accessibility assessments? \textbf{RQ2, Implementation coverage:} which WCAG criteria are instantiated in this framework, and which parts of that implementation are represented by positive evaluation cases? \textbf{RQ3, Detection and cost:} how do the archived worker configurations compare with rule-based checking, batch and per-criterion vision-language assessment, and general-purpose coding agents in agreement with the reference labels and recorded resource use? \textbf{RQ4, Evidence availability:} how do detection patterns and abstentions vary across criteria and model configurations when the available representations and operations differ? The fourth question concerns observed configurations, whose differences in instructions, tools, and inputs are documented in the comparison conditions.

The primary analysis uses \nrows{} page--criterion records from \npages{} pages across \nplatforms{} scholarly platforms, including \npos{} positive and \nneg{} inferred-negative labels. Reference workers recover \agentrecall{} of positives, compared with \axerecall{} for axe-core and \vlmpurerecall{} for an uncued batch vision-language model, at precisions of \agentprecision{}, \axeprecision{}, and \vlmpureprecision{}, respectively. Their higher recall accompanies more findings absent from the reports. All conditions target the saved-page corpus. We pair each cost estimate with its run's predictions, separating the reference workers from later instrumented executions. Differences in tools, instructions, and some models make these descriptive comparisons.

The paper contributes a criterion-specific worker architecture and implementation; an account of translating normative requirements into evidence collection and decision instructions; and an evaluation corpus and reanalysis covering reference-label agreement, abstention, and resource use. Criterion-level comparisons and sensitivities addressing development exposure and report provenance support examination of where browser operations help and where uncertainty remains. These findings motivate a proposed allocation of rule-based, model-based, and interactive assessment within professional auditing, whose effects on auditors' work remain to be evaluated.
\section{Background and Related Work}
\label{sec:related}
\subsection{Accessibility evaluation, conformance, and professional practice}
WCAG organizes requirements under four principles: content should be perceivable, operable, understandable, and robust~\cite{w3cwai}. The criteria differ in what must be inspected and in the context needed to apply them. Some checks concern a single element, while others require considering relationships, alternative presentations, or a sequence of actions. This matters when choosing an evaluation unit. A page--criterion assessment aggregates potentially several element-level observations into one result, whereas a site-level conformance evaluation must also address the scope and representative coverage of the pages and processes examined. We use the former unit to compare recorded predictions and reserve conformance claims for the broader assessment that those predictions do not provide.

WCAG-EM 1.0 describes a structured process for defining evaluation scope, exploring a website, selecting a representative sample, auditing that sample, and reporting findings~\cite{wcagem}. The July 2026 WCAG-EM 2.0 Group Note extends this evaluation framework to digital products and explicitly distinguishes sampled evaluations from whole-website conformance claims~\cite{wcagem2}. These methodologies provide context for professional reports, rather than a guarantee that every unmentioned criterion was exhaustively tested on every page. This distinction is consequential when reusing reports as research data. A positive finding records an identified problem within an audit's scope. Silence may instead reflect sampling, the purpose of a report, or a different page state. Our reference labels preserve the first kind of evidence directly and introduce an additional assumption for the second. We expose that assumption in the dataset description and examine its consequences through sensitivity analyses.

Studies of accessibility evaluation also caution against equating the output of one method with accessibility as a whole. Vigo et al. examine the consequences of sole reliance on automated evaluation tools, including variation in the problems that different tools identify~\cite{vigo2013}. Power et al. show that problems encountered by blind users extend beyond those captured by guideline conformance alone~\cite{power2012}. These findings motivate complementary forms of evaluation and careful interpretation of coverage. A system that detects more criterion-level labels may be useful, but that result does not directly measure the number of people helped, the severity of the barriers removed, or successful completion of a user's task. Those outcomes require evidence beyond the present benchmark.

Professional practice introduces organizational considerations as well as technical ones. Bi et al.'s interviews and survey describe accessibility work across software development settings, including practitioners' experiences with knowledge, resources, and development processes~\cite{bi2022}. Their account situates a detector within a continuing process of identifying, communicating, and addressing problems. For our framework, this motivates recording the evidence behind an assessment in a form that can be examined after execution. A finding can be inexpensive to generate yet costly to verify, particularly when it lacks a clear location or requires reconstructing a transient interface state. We therefore distinguish computational cost from human review cost throughout the evaluation.

\subsection{Rule-based and interaction-based accessibility testing}
Rule-based engines such as axe-core execute checks associated with accessibility requirements~\cite{axecore}, using rendered-document and browser information as well as HTML. Coverage depends on the implemented procedure and inspected state. Our configured axe execution therefore cannot represent every deterministic test or establish that all rule-based methods lack a capability observed in workers.

Dynamic web testing investigates behavior beyond an initial view. Prior systems detect and locate keyboard failures~\cite{keyboard2021}, navigation barriers~\cite{bagel2023}, and dialog-specific navigation failures~\cite{dialog2023}. These are direct predecessors to our interaction tools. We investigate how such operations can be selected and interpreted by workers assessing different named criteria within shared infrastructure.

Mobile accessibility research offers additional precedents for generating and preserving interaction evidence. Groundhog compares interaction modalities while crawling applications and records evidence of accessibility problems~\cite{groundhog2022}. AXNav translates natural-language accessibility tests into execution and replay, supporting inspection of test behavior~\cite{axnav2024}. These systems demonstrate useful roles for executable tests and reviewable records, although mobile application structure, interaction mechanisms, and evaluation scope differ from saved web documents. The relevant comparison is the relationship between a test objective and its evidence, rather than an assumption that results transfer unchanged between platforms. Our framework similarly records operations, but its reporting unit is an assigned WCAG success criterion for a particular page.

TaskAudit is the closest agentic predecessor in this comparison. It generates tasks, executes them through a screen-reader proxy, and analyzes execution records for \emph{functiona11ity errors}, its term for accessibility barriers that emerge through interaction~\cite{zhong2026taskaudit}. The name combines functionality and accessibility; the evaluated barriers concern screen-reader-mediated tasks. Its error taxonomy is related to WCAG. The approaches instead differ in how an assessment begins and what it returns: TaskAudit organizes analysis around generated mobile tasks, while our workers begin with a named criterion and collect evidence relevant to that criterion on a rendered web snapshot. Both approaches require bounding their claims by the operations available to the agent and the behavior their experimental environments support.

\subsection{Language and vision-language models for auditing}
Bassi et al. examine LLM support for HTML validation, accessibility auditing, and generation of accessible snippets, identifying both useful developer assistance and limitations of relying on model judgments~\cite{bassi2025}. GenA11y combines criterion-specific extraction with language-model assessment and explicitly considers differences in the evidence required by static and dynamic checks~\cite{gena11y2025}. Its work establishes that both the criterion and the relevant portion of the page can structure model input. We build on that orientation by giving a worker a criterion-specific set of executable operations and permitting further evidence collection during assessment. Our criterion-level results examine when these additional operations accompany more detected reference cases or more false positives.

AccessGuru combines detection and correction and organizes accessibility problems using syntactic, semantic, and layout distinctions connected to WCAG~\cite{fathallah2025accessguru}. WebAccessVL studies violation-aware vision-language support for web accessibility, including a relationship between a checker and repair~\cite{li2026webaccessvl}. These systems distinguish finding a violation, locating the affected element, and correcting it. Successful repair under a supplied violation signal is a different outcome from independently identifying whether an unlabelled page violates a criterion. We compare recorded detection outputs in this study and do not evaluate the quality, side effects, or maintainability of repairs. Similarly, our page--criterion scoring does not establish whether a generated description accurately localizes every issue it mentions.

ScreenAudit uses language models with mobile interface information and screen-reader-related evidence to identify accessibility errors~\cite{screenaudit2025}. Alongside TaskAudit, it illustrates how the model's input representation and the method used to gather that representation shape the problems available for analysis. In our vision-language conditions, the model receives a screenshot and accessibility-tree information assembled by the evaluation pipeline. The accessibility tree is the browser's representation of relevant roles, names, states, and relationships exposed for assistive technologies. It adds information that pixels alone do not contain, while remaining distinct from an observed sequence of keyboard or screen-reader interactions. Consequently, we describe these conditions by their supplied evidence rather than treating them as image-only assessment.

Gu et al.'s AAA framework addresses scalability through page sampling and multimodal assistance to human auditors~\cite{gu2025scalable}. It supports both the selection of pages to inspect and the auditor's assessment of those pages. Sampling pages and allocating assessment methods across criteria are related but distinct decisions. A system could use a sampling method to select pages and a criterion-specific policy to select operations within them, although the performance of that combination would require evaluation. We use a fixed archival sample and therefore leave adaptive page selection outside our empirical claims.

\subsection{Agent architecture and human oversight}
The framework's alternating reasoning and tool use follows the general approach exemplified by ReAct~\cite{react2023}. A model receives observations, selects an action, and uses the returned evidence in a subsequent decision. Adapting that approach to accessibility requires specifying more than task completion: the worker must connect an observed property to the conditions of a normative requirement. We implement this through shared execution infrastructure and criterion-specific instructions. The design makes the instructions and tool allocation inspectable. Our evaluation examines this configured architecture, with tool access and criterion-specific guidance changing together relative to some comparison conditions.

Human--AI interaction research provides a basis for discussing how such outputs might be used. Amershi et al. identify design considerations for communicating system capabilities and supporting correction~\cite{amershi2019}. Bu\c{c}inca et al. examine overreliance and show why presenting an explanation alone should not be assumed to produce appropriate reliance~\cite{bucinca2021}. Bansal et al. likewise investigate the conditions for complementary human--AI performance~\cite{bansal2021}. These studies motivate distinguishing an inspectable record from a demonstrated benefit to decision making. Our evaluation measures automated predictions and their recorded execution, while the Discussion uses this literature to formulate testable implications for auditor review.

\begin{table}[t]
\centering
\small
\caption{Selected predecessors and their assessment organization. All descriptions concern the cited systems, whose datasets and metrics differ from ours; the table is not a performance ranking.}
\label{tab:prior}
\begin{tabularx}{\linewidth}{p{2.1cm}XX}
\toprule
System & Evidence and execution & Assessment organization \\
\midrule
GenA11y~\cite{gena11y2025} & Criterion-relevant elements extracted from web pages & Criterion-specific model detection \\
AccessGuru~\cite{fathallah2025accessguru} & Web-content representations and model-based analysis & Detection and correction; categories connected to WCAG \\
WebAccessVL~\cite{li2026webaccessvl} & HTML, rendering, and checker-provided violations & Violation-conditioned repair and refinement \\
BAGEL~\cite{bagel2023} & Automated keyboard navigation on web interfaces & Navigation-based accessibility barriers \\
Groundhog~\cite{groundhog2022} & Mobile crawling and comparison across interaction modalities & Accessibility failures with execution evidence \\
AXNav~\cite{axnav2024} & Natural-language tests translated into mobile execution & Test replay and inspection \\
ScreenAudit~\cite{screenaudit2025} & Mobile interface and screen-reader-related evidence & Model-based screen-reader error detection \\
TaskAudit~\cite{zhong2026taskaudit} & Generated mobile tasks executed through a screen-reader proxy & Functiona11ity error categories connected to WCAG \\
This framework & Criterion-specific tools operating rendered web snapshots & Page--criterion assessments with execution records \\
\bottomrule
\end{tabularx}
\end{table}

\section{The Worker-Agent Framework}
\label{sec:framework}
\subsection{From a success criterion to an executable assessment}
\begin{figure}[t]
\centering
\includegraphics[width=\linewidth]{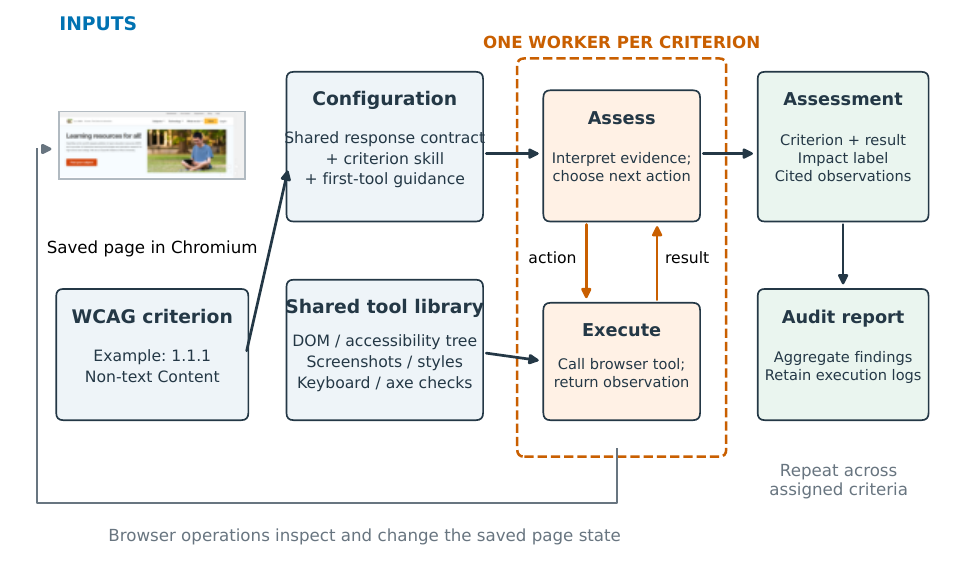}
\caption{Criterion-specific assessment within shared execution infrastructure. Tool results enter the next model step; the final result retains its criterion and supporting execution record. A saved document may support browser operations without reproducing the original production service.}
\Description{The target snapshot and criterion enter a worker configuration containing its skill and declared tools. A loop alternates model assessment, tool execution, and returned observations. It ends with an assessment, impact label, and evidence record. An authentic saved-page screenshot illustrates the input. Shared infrastructure comprises the browser, tool implementations, response contract, and logging.}
\label{fig:framework}
\end{figure}

The framework takes a target document and a WCAG success criterion and returns an assessment with a record of the evidence collected. Its purpose is to share the infrastructure needed to conduct an audit while retaining the distinctions between individual requirements. The same browser query may support several criteria, but the inference drawn from its result can differ. An element's accessible name, for example, may be relevant to both non-text content and name--role--value requirements. Assigning a worker to one criterion gives the decision an explicit scope and permits the associated instructions to state what additional evidence is needed before reporting a finding. Figure~\ref{fig:framework} summarizes this relationship.

We operationalize each criterion by specifying the properties to inspect, the operations through which they can be observed, and the conditions under which those observations support a result. Operationalization here means translating a normative requirement into an executable assessment procedure; it does not replace or modify the requirement. A worker's \emph{skill document} is a plain-text instruction file, written in Markdown, supplied to the model for one success criterion. The runtime supplies this text as instructions governing the model's behavior and appends the shared output instructions. It specifies what to inspect, relevant exceptions, and how to interpret tool results; it is an instruction document rather than executable browser code. Shared instructions establish the response format, while a worker configuration identifies its declared tools and, where specified, the required first operation. The framework thus separates reusable execution mechanisms from instructions whose content depends on the criterion.

This decomposition separates maintenance responsibilities: corrections to a browser operation apply to all workers using it, while criterion-specific clarifications belong in the worker's instructions. Shared defects can affect multiple workers and pages, linking their outcomes despite the modular organization. The development-exposure analysis accounts for this possibility by excluding whole pages.

\subsection{Shared execution and criterion-specific tools}
The worker loop alternates reasoning and action~\cite{react2023}: the model interprets previous observations, requests an operation, and receives its result before deciding whether to continue. A shared base class coordinates headless Chromium and browser-facing tools. Its current default limits execution to 30 iterations to prevent indefinite runs. Reaching that bound is an execution limitation, not evidence of conformance; errors remain separate from completed positive and negative assessments.

The shared tool library supports document and accessibility-tree extraction, element queries, screenshots, keyboard and pointer operations, computed-style inspection, contrast measurement, and axe-based checking. The Document Object Model (DOM) is the browser's structured representation of the document and its elements. DOM queries can retrieve attributes and relationships, while computed styles and screenshots supply information about their rendered presentation. Keyboard operations add observations about state changes and focus movement. Combining these sources is useful because an attribute can suggest a problem without establishing its visible or behavioral consequence. The worker's instructions determine which combination is relevant to the assigned criterion. Table~\ref{tab:tools} summarizes the library by the evidence each group produces. The supplementary tool inventory lists all 31 registered operations, their parameters, and implementation files; the complete server and browser-tool source is supplied in \texttt{tool\_library/}. The supplementary \texttt{worker\_skills/} directory contains all 40 criterion instruction files and the shared response instructions. Batch, per-criterion, and coding-agent prompt assets are supplied in \texttt{comparison\_prompts/}.

\begin{table}[t]
\caption{Tool groups and the evidence they return. These are implementation capabilities, not claims of complete criterion coverage. The supplementary tool inventory lists every registered operation and its parameters.}
\label{tab:tools}
\small
\begin{tabularx}{\linewidth}{p{.23\linewidth}X}
\toprule
Tool group & Operations and returned evidence \\
\midrule
Document inspection & Extract document structure and accessibility information; query elements and attributes; inspect computed styles and page readiness. \\
Visual inspection & Capture screenshots or navigation video; measure contrast; test resizing, reflow, text spacing, and color presentation. \\
Interaction and focus & Send keyboard, pointer, and viewport commands; record focus sequences; activate widgets; inspect focus indicators and reading order. \\
Rule-based checks & Run axe-core on a document or accessible frames; inspect ARIA attributes, control names, media, language, animations, and PDF properties. \\
Alternative access and reporting & Inspect or control a browser through an external browser-control service; analyze public URLs through an external model service; simulate tree traversal and aggregate tool results. \\
\bottomrule
\end{tabularx}
\end{table}

Tool selection is specialized through the worker configuration. For example, the Contrast (Minimum) worker requests rendered-contrast assessment first and can supplement it with style queries, screenshots, and other checks. The No Keyboard Trap worker starts with the \emph{widget probe}, a scripted operation that activates controls such as tabs, radio buttons, and dropdown controls and records the focus sequence produced by keyboard navigation. Repeated focus on a control is reported as a candidate problem, which still requires interpretation against the criterion. The worker can request further keyboard-navigation and interaction operations to investigate that candidate. The first-operation instruction provides a starting point grounded in the kind of evidence the criterion requires. The \emph{tool server} is the software service that exposes these browser operations through named calls using the Model Context Protocol (MCP). It receives an operation name and arguments, executes the corresponding implementation, and returns an observation to the worker. The runtime can also call server tools outside the worker's declared list; the log marks these calls as non-included, meaning that they were not in that list. The declared list therefore guides tool selection; it is not an enforced capability boundary. A tool's structured summary is itself a product of an implemented procedure, so the analysis of a disputed finding may require examining the underlying operations rather than treating the summary as independently verified evidence.

Workers can reuse deterministic checks: an axe result can identify candidates or show that particular checks returned no findings, after which the worker can seek further evidence. This avoids asking a model to reproduce every low-level calculation. The evaluation therefore concerns the combined behavior of model decisions and deterministic tools.

The shared response contract requests true, false, or error, a summary, and any structured findings. True reports a violation of the assigned criterion; false reports none found through the worker's procedure; error records incomplete execution. These results describe the retained assessment operations rather than certify conformance. The per-criterion VLM's undetermined response is a separate protocol feature, not a terminal category implemented identically by every method.

The archived \texttt{confidence} field describes assessed impact through labels such as High (Major) and Medium (Moderate), not calibrated probabilities of correctness. We retain it but do not weight predictions or select probability thresholds from it. Impact, uncertainty about correctness, and execution status differ: a potentially severe issue can have weak evidence, whereas an error means the assessment did not finish.

\subsection{Authoring and implementation coverage}
To author a worker, we identify the requirement, select evidence-producing operations, write decision guidance, inspect example executions, and revise instructions or tools in response to problems. We did not compare authoring methods or systematically measure authoring time. The contribution is the configuration structure and instantiated criteria, with records connecting instructions to behavior.

For Non-text Content (SC 1.1.1), the requirement is that a text alternative convey the purpose of non-text content, subject to the criterion's exceptions~\cite{nontext}. We instantiate the image-assessment portion by linking three decisions to observable evidence. First, the worker locates images and image-based controls, including those in accessible embedded documents. Second, it inspects their accessible names, nearby text, and visual context to distinguish information-bearing content from decoration. Third, it reports a finding when the observed alternative fails to express the image's information or the control's function, rather than treating every empty alternative as a violation. The instruction file names the relevant checks and exceptions; the configuration makes the corresponding tools available. The supplementary appendix, ``Complete authoring example: Non-text Content,'' provides the actual instruction excerpts, tool mapping, decision cases, and archived checks. It also identifies aspects of the broader criterion that the image-focused implementation does not explicitly cover.

Decision guidance must explain when a suspicious pattern does not establish a violation. An empty alternative-text attribute may be appropriate for decoration, but inadequate for an informative image~\cite{nontext}. Likewise, repeated focus requires checking appropriate exit actions and legitimate widget behavior before reporting a keyboard trap~\cite{keyboardtrap,radiopattern}. Such cases require contextual evidence; criterion text, tool output, and verdict express different parts of the assessment.

The implementation contains \nagents{} workers. Of these, \ncovered{} correspond to WCAG 2.1 Level A and AA success criteria, and one addresses WCAG 2.2 criterion 2.5.7, Dragging Movements~\cite{wcag22}. The eleven WCAG 2.1 A/AA criteria without workers are 2.4.5, 2.4.7, 2.5.2, 2.5.4, 3.2.1--3.2.4, 3.3.1, 3.3.3, and 4.1.3. We report this as an implementation inventory. The records do not support assigning each omission to a fundamental limit of automation or to a specific tooling deficiency. Such a classification would require a separate analysis of the evidence and procedures that could address each requirement.

The primary benchmark evaluates all 40 implemented criteria but contains positive reference cases for only 15. For the other 25 criteria, the available rows can reveal disagreement with inferred-negative labels, but cannot estimate recall. This includes the WCAG 2.2 worker, whose five records have inferred-negative labels. Its inclusion reflects the report-based construction described in Section~\ref{sec:benchmark}, rather than an attempt to represent all requirements added in WCAG 2.2. This reporting separates the ability to instantiate a worker from evidence that the worker detects violations of its criterion.

\subsection{Execution evidence and a worked example}
The framework records tool names, arguments, observations, timestamps, and final results, distinguishing requested operations from returned evidence and the worker's interpretation. To inspect a claim, a reviewer must locate its supporting observation and check whether its scope and fidelity justify the interpretation. A model-generated explanation alone can overstate what was established.

\begin{figure}[t]
\centering
\includegraphics[width=\linewidth]{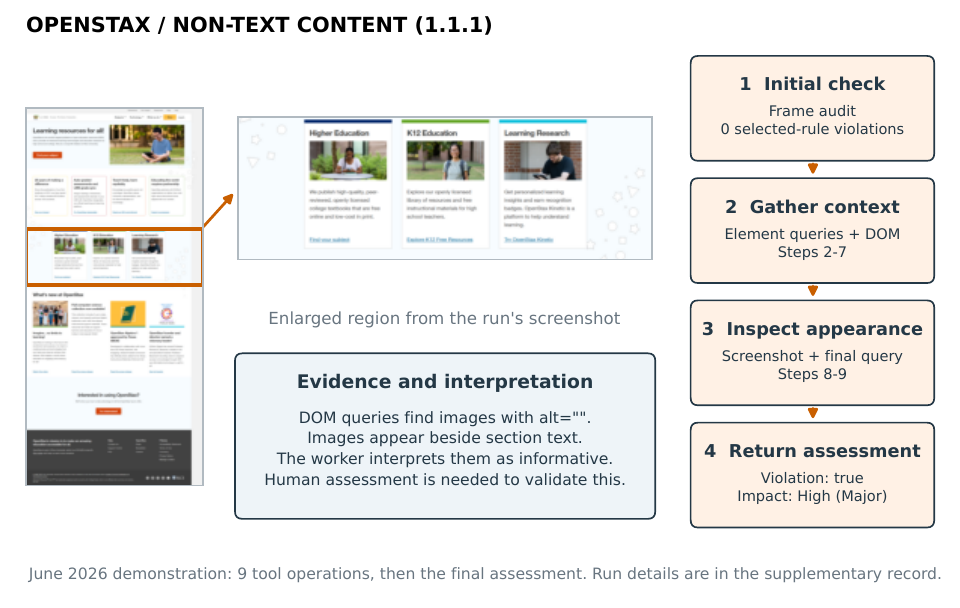}
\caption{Execution example from a June 2026 worker assessment of OpenStax Non-text Content. The screenshot was captured by this run. Numbered groups follow the nine tool operations and terminal decision; excerpts summarize observed evidence. The worker interpreted the empty-alt images as informative. Its positive criterion-level score does not independently validate that interpretation.}
\Description{A screenshot overview locates the Higher Education, K12 Education, and Learning Research cards. An enlarged crop makes their images and nearby text legible. Four numbered groups show the initial frame audit, DOM and element queries, screenshot and final query, then the positive decision. The arrows connect observations to the worker decision, not to an independent issue-level validation.}
\label{fig:trace}
\end{figure}

A June 2026 assessment of Non-text Content on the OpenStax landing-page snapshot illustrates the sequence (Figure~\ref{fig:trace}). The worker first requested an axe-based frame audit: an operation that runs selected axe-core rules on the main document and embedded documents that the browser permits it to inspect from the same origin (the same protocol, host, and port). An embedded document, or frame, is a page displayed inside another page. This operation returned no violations for the selected checks. It then queried image and related elements, extracted the DOM, and gathered additional attributes and context through further element queries. After capturing a screenshot, it made one final query before returning its assessment. The archived screenshot in the figure comes from that execution, allowing the displayed page region to be linked directly to the recorded operation. Its final assessment identified empty-alt images that it interpreted as informative, while excluding other empty-alt images it interpreted as decorative. The sequence shows the worker seeking contextual evidence after an initial deterministic check returned no finding.

The OpenStax report and worker both return positive Non-text Content assessments, so the page--criterion row scores as a true positive. This score does not require identifying the same image as the report: criterion-level agreement can coexist with different issue locations or explanations. The example documents execution and scoring, without independently validating an additional violation. The supplementary \texttt{worked\_trace.json} identifies the run and operations.

\section{The LAA Evaluation Benchmark}
\label{sec:benchmark}
\subsection{Reference labels and sampling}
We constructed the evaluation records from accessibility reports commissioned by the Library Accessibility Alliance for scholarly platforms~\cite{laa2024}. A record pairs a page with a targeted success criterion. A positive label indicates that the source material records a finding associated with that page and criterion; a negative label is inferred when no corresponding finding is listed. We refer to these as \emph{reference labels} to distinguish the report-derived representation from an exhaustive assessment of the saved page. This distinction is especially important for negative labels, which were not collected through an independent pass/fail audit of every sampled page--criterion pair.

We constructed 294 records from positives and five sampled negatives per criterion, using seed 42. We excluded ACS Publications because its website had changed by SingleFile capture time and the report's page versions were unavailable, removing 44 records across four pages: 16 positive and 28 inferred negative. This avoided evaluating labels against known replacement pages. The remaining 250 records cover 24 pages from 11 scholarly platforms, each linked to a self-contained HTML capture configured for a 1280-by-720 primary viewport at the initial scroll position. This content-heavy corpus reflects available reports and intended captures, not random sampling of the web.

The primary population retains all \nrows{} records: \npos{} positive labels and \nneg{} inferred negatives across 40 criteria. Thirty-nine criteria belong to WCAG 2.1 Level A or AA; the remaining criterion is WCAG 2.2 criterion 2.5.7, Dragging Movements. Its inclusion arose during construction against the LAA reports rather than from a separate sampling of WCAG 2.2 requirements. We prioritized recent reports to reduce the interval in which the evaluated webpages might have changed, and this criterion appeared in that report material. The five sampled records for it have inferred-negative labels. We retain them to preserve the report-derived benchmark, without claiming systematic coverage of WCAG 2.2. A supplementary analysis restricts the population to the 245 WCAG 2.1 records. The supplementary document and supporting records are provided as ancillary files accompanying this preprint.

Sampling increases the share of positive cases to support analysis of missed findings; that share does not estimate violation prevalence in deployment. Precision depends on that distribution as well as on the validity of the inferred negatives. We therefore interpret it as agreement with this labelled sample rather than as a predicted proportion of correct findings on arbitrary websites.

The source reports also differ in provenance. Two vendors, Deque and Accessiblu, account for the primary rows, and some findings in the Deque reports are marked as identified through axe, either alone or together with manual examination. Such findings are not independent of the rule engine used in our static condition. We retain them in the primary population because they are part of the source reports, and separately exclude all rows whose provenance includes axe. Vendor-specific analyses use each vendor's own rows and positive denominator. These analyses make the effects of source composition visible without treating a vendor as a randomly assigned experimental condition.

\subsection{Saved documents and experimental provenance}
A \emph{rendered snapshot} in this study is a saved HTML document opened in a browser, with the associated content available in that capture. Rendering permits inspection of the DOM, accessibility tree, appearance, and some interactions. It does not imply that the saved document reproduces the production service's authentication, remote requests, timers, or state transitions. The original audit may also have examined a different version of the service. Agreement between an automated finding and a report label therefore involves both a criterion-level mapping and an assumption that relevant evidence remains available in the captured document. We retain rows flagged as requiring unavailable behavior in the primary analysis and examine their exclusion separately.

SingleFile embeds page resources into self-contained HTML documents~\cite{singlefile}, reducing dependence on production resources during offline evaluation. All conditions use this captured corpus, but their tools and prompts determine the observations extracted. Reproducibility records link each document and success criterion to individual runs.

The reference worker predictions carry the archived failure-analysis flags; subsequent instrumented runs additionally record usage. These are separate executions on the same corpus. We preserve their prediction hashes, model metadata, available configurations, and snapshot mappings to avoid transferring annotations or costs between runs. Missing information, including matching reference-worker usage, remains explicit.

\subsection{Development exposure and sensitivity populations}
The archive includes diagnostic and verification datasets used after changes to workers or tools. Their union touches \ndevrows{} of the 250 evaluation rows and spans \ndevpages{} of the 24 pages. Because changes can affect shared operations and criterion-specific instructions, a row need not appear in a diagnostic file to be influenced by development on another row from the same page. We therefore do not treat the remaining 223 rows as an independent held-out set. Nor does use of a different model during a diagnostic pass remove the possibility that an instruction change benefits subsequent evaluation.

We calculate a stricter descriptive sensitivity by excluding every page represented in those diagnostic datasets. On the full population, this leaves \nheldrows{} rows on five pages, including \nheldpos{} positives. The smaller set examines whether the observed pattern is confined to documented development pages; it does not establish prospective generalization or erase other possible development exposure. Additional sensitivities separate vendors, exclude axe-associated reference findings, exclude rows flagged as requiring unavailable dynamic behavior, and omit each platform in turn. Each comparison uses the same population for every condition being compared and reports its positive denominator.

\section{Comparison Conditions}
\label{sec:conditions}
\subsection{Rule-based and vision-language assessment}
The rule-based condition uses axe-core with WCAG 2 A/AA and WCAG 2.1 A/AA tags~\cite{axecore}. We use the saved output of that axe-core execution, which records a decision for each page--criterion pair and rule identifiers for detected issues. During preparation of the analysis, we found that a separate summary spreadsheet contained different axe predictions for 12 of the 250 records. We therefore calculated the reported metrics from the execution output, matching records by identifier and checking that their reference labels agreed. This was a correction to the source used for scoring, not a new axe run or a change to the reference labels. The supplementary provenance record identifies the selected output and its file hash. This condition establishes the observed performance of the configured engine run; it does not measure every check that a deterministic system could implement or every state that an interactive test could reach.

The batch vision-language protocol uses three model calls for each page. Stage 1 assesses Perceivable requirements (principle 1, concerning presentation of content) and Robust requirements (principle 4, concerning compatibility with assistive technologies). Stage 2 assesses Operable requirements (principle 2, concerning controls and navigation), and stage 3 assesses Understandable requirements (principle 3, concerning comprehensible content and operation). Each prompt lists the corresponding WCAG 2.1 Level A and AA criteria. The calls share the screenshot and accessibility-tree representation; reported criterion identifiers are then matched to the page's evaluation rows.

The \emph{uncued} condition receives those representations without additional hints. The \emph{cued} condition also receives a text block headed ``STATIC ANALYSIS HINTS,'' containing criterion-specific counts produced by simple rules over the captured element data. We call these rules \emph{heuristics} because they identify patterns worth inspecting, rather than establish violations. For example, the image rule selects elements whose tag is \texttt{img} and whose captured visibility flag is true, then counts those with an empty accessible name after whitespace is removed. Its prompt line has the form ``1.1.1: N potential issue(s) detected,'' where N is that count. An accessible name is the text used to identify an element to assistive technology; an empty name can be appropriate for a decorative image, so the count alone is not a criterion verdict. These counts are computed by our preprocessing code, not taken from axe-core results or reference labels.

In each cued batch call, the hint block contains counts for criteria belonging to that stage's principle groups. In the per-criterion condition, it contains only the count for the assigned criterion, and is omitted when that count is zero or unavailable. Thus, the same type of cue is supplied at different scopes. The supplementary \texttt{comparison\_prompts/} directory contains the three batch templates, per-criterion template, and cue-generation code so that both the instructions and the generation rule can be inspected.

The per-criterion VLM condition requests one assessment for each page--criterion pair. It includes the criterion's normative statement, the page representation, a criterion-specific heuristic cue, and an explicit option to return an undetermined result. We use the corrected, neutral-abstention variant in the archive; its instruction does not name categories on which the model should abstain. An \emph{abstention} is an explicit refusal to decide whether the criterion is violated on the supplied evidence. It is distinct from a negative assessment and from an execution error. An earlier variant is retained in the archive but does not define the reported condition. The per-criterion protocol changes task organization and output behavior, while the worker comparison additionally changes tools and criterion-specific guidance, so neither contrast is interpreted as a complete causal isolation of one factor.

\subsection{Worker and general-purpose coding-agent configurations}
We analyze reference worker predictions attributed in the archive to GPT-5.5 and separate September 2026 runs recorded with Gemini 3.1 Pro, Claude Sonnet 5, and GPT-5.5. A second complete September 2026 GPT-5.5 pass permits a limited repeatability check. Each worker receives its assigned criterion and can gather evidence through its configured tools. The instrumented runs include usage records and manifests identifying the evaluated snapshot corpus. Model names are reported as recorded in the metadata, and the associated usage is not transferred to the historical predictions. Additional archived batch VLM runs provide a descriptive check across seven recorded model configurations.

General-purpose coding agents provide a comparison with tools that can inspect saved source files through their own execution environments. We analyze Claude Code and Codex under two instructions: a brief audit request, labelled \emph{naive} in the archive, and a \emph{parity} request enumerating the targeted criteria. The latter label refers to criterion enumeration, not identical system prompts, tools, or evidence processing. Claude Code records Claude Opus 5, while Codex records GPT-5.5. Each retains its built-in system instructions. Consequently, differences between these products cannot be attributed solely to a model, a prompt, or a particular tool-use strategy.

The coding-agent metadata documents isolation controls: work directories outside the repository, content-hashed directory names, disabled user configuration and external integrations, and restrictions on network tools. Shell and local file operations remain available. These controls reduce access to project-specific rubrics and evaluation context, but also define a narrower environment than an unrestricted developer session. We describe the condition accordingly, without assuming that the brief prompt represents every developer's normal request. The source expert audit is used as a reference, not as a newly executed human comparison condition; no human auditing time or performance is measured here.

\subsection{Metrics and resource accounting}
We calculate micro-averaged precision as the proportion of positive predictions with positive reference labels. Recall divides true positives by all reference-positive rows in the evaluated population. An abstention or error therefore does not recover a positive label, although it remains separately identified rather than being relabelled a negative decision. F1 is the harmonic mean of this precision and recall. The confusion tables show positive and negative decisions separately from abstentions and errors, allowing readers to distinguish a low detection yield from confident incorrect decisions. Undefined ratios are reported as unavailable, including recall for a criterion without positive cases.

We additionally calculate conditional recall on rows where the per-criterion VLM commits to a positive or negative decision, and apply that same subset to the other conditions. This makes its selection effect visible: conditional recall addresses performance after abstaining, whereas full-population recall addresses how many reference findings the entire procedure recovers. For costs, we use complete 250-row workload totals, matching the primary population. Page-level calls are shared across criteria, so per-row averages describe an allocation of total usage rather than a separate billed request for every criterion. Costs are paired with accuracy from that same run and population. Logged token estimates, API-equivalent subscription estimates, summed processing spans, and elapsed batch time are not treated as interchangeable quantities.

OpenAI Codex assisted development of the deterministic reanalysis scripts and vector-figure code used for this revision. The scripts operate on archived predictions, labels, metadata, and execution records; they do not request new model judgments or generate replacement reference labels. The accompanying source and row-level outputs expose the joins, population filters, metric definitions, and figure inputs for inspection. AI assistance with these research artifacts is distinguished from the model configurations evaluated as audit conditions.

\section{Evaluation and Findings}
\label{sec:findings}
\subsection{Overall agreement and implementation coverage}
RQ1 and RQ2 concern the shared execution architecture and its 39 WCAG 2.1 A/AA workers plus one WCAG 2.2 worker. All are represented in the benchmark, but only 15 criteria have positive detection cases. Table~\ref{tab:primary} addresses detection and cost (RQ3); criterion-level results and abstentions address evidence availability (RQ4). Results are descriptive: rows share page content and capture conditions, and pages share platform conventions, so rows are not independent experimental replications.

\begin{table}[t]
\centering
\small
\caption{Agreement with reference labels on the full 250-row benchmark (78 positives, 172 inferred negatives). U denotes abstention; E denotes execution error. Recall includes every reference-positive row in its denominator.}
\label{tab:primary}
\begin{tabular}{lrrrrrrrrr}
\toprule
Condition & TP & FP & TN & FN & U & E & P & R & F1 \\
\midrule
axe-core & 28 & 3 & 169 & 50 & 0 & 0 & 0.90 & 0.36 & 0.51 \\
Batch VLM, uncued & 52 & 33 & 139 & 26 & 0 & 0 & 0.61 & 0.67 & 0.64 \\
Batch VLM, cued & 58 & 33 & 139 & 20 & 0 & 0 & 0.64 & 0.74 & 0.69 \\
Per-criterion VLM & 46 & 28 & 79 & 5 & 92 & 0 & 0.62 & 0.59 & 0.61 \\
Workers, reference run & 67 & 53 & 119 & 11 & 0 & 0 & 0.56 & 0.86 & 0.68 \\
Claude Code, naive & 47 & 26 & 146 & 31 & 0 & 0 & 0.64 & 0.60 & 0.62 \\
Claude Code, parity & 60 & 49 & 123 & 18 & 0 & 0 & 0.55 & 0.77 & 0.64 \\
Codex, naive & 25 & 10 & 162 & 53 & 0 & 0 & 0.71 & 0.32 & 0.44 \\
Codex, parity & 42 & 24 & 148 & 36 & 0 & 0 & 0.64 & 0.54 & 0.58 \\
\bottomrule
\end{tabular}
\end{table}

Workers in the reference run recover \agentrecall{} of the reference positives at precision \agentprecision{}. Axe-core has higher precision, \axeprecision{}, with recall \axerecall{}. The uncued batch VLM falls between them in recall, \vlmpurerecall{}, and returns precision \vlmpureprecision{}. Thus, the worker configuration identifies more positive-labelled rows while also returning more findings absent from the reports. This tradeoff is visible in F1: the workers in the reference run score \agentfone{}, the uncued VLM \vlmpurefone{}, and the cued VLM \vlmcuedfone{}. F1 reflects both missed reference cases and positive predictions absent from the reports.

Enumerating criteria increases Claude Code's recall from \claudeNaiveR{} to \claudeParityR{} and Codex's from \codexNaiveR{} to \codexParityR{}, with additional false positives in both cases. Thus, audit instructions matter in these configurations. Criterion enumeration does not equalize system instructions, tools, or models: Claude Code uses a different recorded model from the workers. These conditions illustrate general-purpose audit behavior under the recorded instructions.

\begin{figure}[t]
\centering
\includegraphics[width=\linewidth]{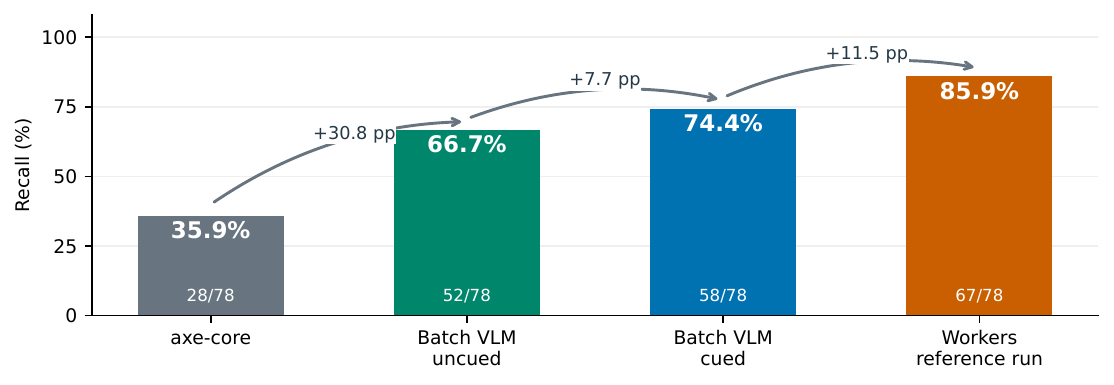}
\caption{Recall on the full 250-row benchmark, with 78 positive reference labels. Arrows show observed percentage-point differences between configurations; they do not describe cumulative stages of an evaluated workflow.}
\Description{Four bars show axe-core at 28 of 78 positives, uncued batch VLM at 52, cued batch VLM at 58, and reference workers at 67. Arrows mark differences of 30.8, 7.7, and 11.5 percentage points.}
\label{fig:metrics}
\end{figure}

\subsection{Criterion-level variation and abstention}
Aggregate recall conceals variation in both sample size and detection pattern. Info and Relationships and Name, Role, Value each contribute 17 positive rows, together accounting for 34 of the 78 positives. Several other criteria contribute only one. Table~\ref{tab:criteria} reports all positive denominators so that a result on one case is not visually equated with a result on seventeen. The table also shows that additional browser operations do not produce uniformly higher detection counts. For Non-text Content, the per-criterion VLM recovers seven of eight reference cases, compared with six for the workers in the reference run. For Contrast (Minimum), the uncued VLM recovers four of five, compared with three for the workers. These observations support choosing assessment methods with attention to the criterion and available evidence.

\begin{table}[t]
\centering
\small
\caption{Detection counts for all criteria with positive reference cases. Cells give true-positive counts beside the reference-positive denominator; no detection-coverage inference is made for criteria without positives.}
\label{tab:criteria}
\begin{tabular}{lrrrrrr}
\toprule
Criterion & Positive & axe & Uncued & Cued & Per-crit. & Workers \\
\midrule
1.1.1 & 8 & 1 & 4 & 6 & 7 & 6 \\
1.3.1 & 17 & 8 & 14 & 17 & 10 & 14 \\
1.3.2 & 1 & 0 & 0 & 0 & 0 & 1 \\
1.4.1 & 1 & 0 & 0 & 0 & 0 & 0 \\
1.4.3 & 5 & 1 & 4 & 2 & 4 & 3 \\
1.4.11 & 1 & 0 & 1 & 1 & 1 & 1 \\
2.1.1 & 8 & 0 & 0 & 2 & 0 & 7 \\
2.1.2 & 2 & 0 & 0 & 0 & 0 & 2 \\
2.2.2 & 1 & 0 & 1 & 1 & 0 & 1 \\
2.4.3 & 7 & 0 & 4 & 4 & 3 & 6 \\
2.4.4 & 4 & 2 & 3 & 3 & 2 & 4 \\
2.4.6 & 3 & 0 & 1 & 2 & 3 & 3 \\
2.5.3 & 1 & 0 & 1 & 1 & 1 & 0 \\
3.3.2 & 2 & 0 & 2 & 2 & 0 & 2 \\
4.1.2 & 17 & 16 & 17 & 17 & 15 & 17 \\
\bottomrule
\end{tabular}
\end{table}

Keyboard-related results show a different pattern. Workers in the reference run recover seven of eight Keyboard cases and both No Keyboard Trap cases; the uncued batch VLM recovers none of these cases. For Focus Order, however, the uncued VLM recovers four of seven cases and the workers six. The Focus Order result shows that noninteractive representations can support detection on some behavior-related cases: source and layout evidence may expose aspects of a requirement without reproducing its complete behavior. The workers' advantage is most pronounced on the Keyboard and No Keyboard Trap cases in this archive; interaction, instructions, and tools vary together in this comparison.

\begin{figure}[t]
\centering
\includegraphics[width=\linewidth]{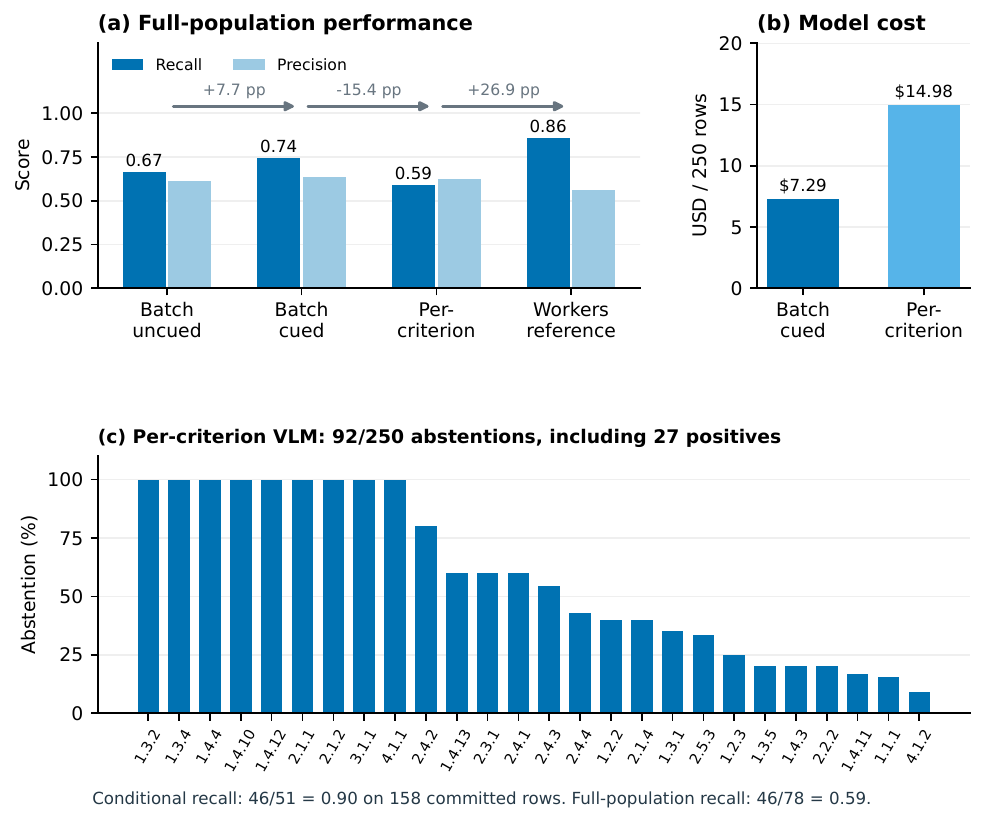}
\caption{Prompting configurations, resource use, and abstention on the full benchmark. The per-criterion condition costs more than the cued batch condition and abstains on 92 records. Its conditional recall uses only committed rows, as stated beneath the panels. Bars summarize recorded configurations rather than isolated effects of granularity or interaction.}
\Description{The first panel compares precision and full-population recall for uncued batch, cued batch, per-criterion VLM, and reference workers. The second shows costs of 7.29 and 14.98 USD for cued batch and per-criterion processing. The lower panel shows abstention rates by criterion, with all positive-denominator comparisons retaining 78 reference positives.}
\label{fig:granularity}
\end{figure}

The per-criterion VLM explicitly abstains on \abstainn{} of \nrows{} rows, or \pcAbstainPct{}\%, including 27 positive-labelled rows. Its recall over the full positive population is \pcrecall{}, whereas its conditional recall on committed rows is 0.90. The difference comes from the denominator: the committed population contains \committedn{} rows and 51 positives, of which the model recovers 46. On precisely that same subset, the cued batch VLM recovers 45 positives, and the uncued batch VLM and workers in the reference run each recover 42. These values describe the rows selected by the per-criterion procedure, not a performance advantage over methods evaluated on all 78 positives. Figure~\ref{fig:granularity} places these two denominators alongside the prompting conditions and their costs. The supplementary analysis preserves the matching-subset comparisons and abstention counts by criterion.

Abstention varies by criterion: the per-criterion VLM declines all seven No Keyboard Trap rows, including two positives, and eleven Keyboard rows, but commits on some other behavior-related cases, including Focus Order. Neither abstention nor commitment independently establishes whether the input contains sufficient evidence. Auditor-facing systems should therefore expose the reasons for declining and the evidence supporting committed decisions, rather than treating either response as inherently reliable.

\subsection{Variation across recorded model configurations}
The seven archived batch VLM configurations range in primary recall from 0.47 to 0.77 (Table~\ref{tab:models}). Differences in overall performance do not translate into a uniform recovery of the keyboard cases. None identifies either No Keyboard Trap positive, while Keyboard true positives range from zero to four of eight. Focus Order has nonzero detections in several configurations. The two No Keyboard Trap cases are therefore consistently missed in these configurations, while detection of other keyboard and focus issues varies by criterion and model. This consistent gap motivates testing which additional observations help assess these checks.

\begin{table}[t]
\centering
\small
\caption{Archived batch VLM configurations on the primary 250 rows. These runs provide a descriptive model check, not an isolated manipulation of model capability.}
\label{tab:models}
\begin{tabular}{lrrrrrrr}
\toprule
Recorded model & TP & FP & U & E & P & R & F1 \\
\midrule
GPT-5.5 & 58 & 33 & 0 & 0 & 0.64 & 0.74 & 0.69 \\
Claude Sonnet 5 & 60 & 50 & 0 & 0 & 0.55 & 0.77 & 0.64 \\
Gemini 3.5 Flash & 48 & 24 & 0 & 0 & 0.67 & 0.62 & 0.64 \\
Kimi K2.6 & 52 & 22 & 0 & 0 & 0.70 & 0.67 & 0.68 \\
Gemma 4 (31B) & 51 & 26 & 0 & 0 & 0.66 & 0.65 & 0.66 \\
GLM-5.2 & 46 & 20 & 0 & 0 & 0.70 & 0.59 & 0.64 \\
Qwen3.5 (397B) & 37 & 25 & 0 & 0 & 0.60 & 0.47 & 0.53 \\
\bottomrule
\end{tabular}
\end{table}

\begin{figure}[t]
\centering
\includegraphics[width=\linewidth]{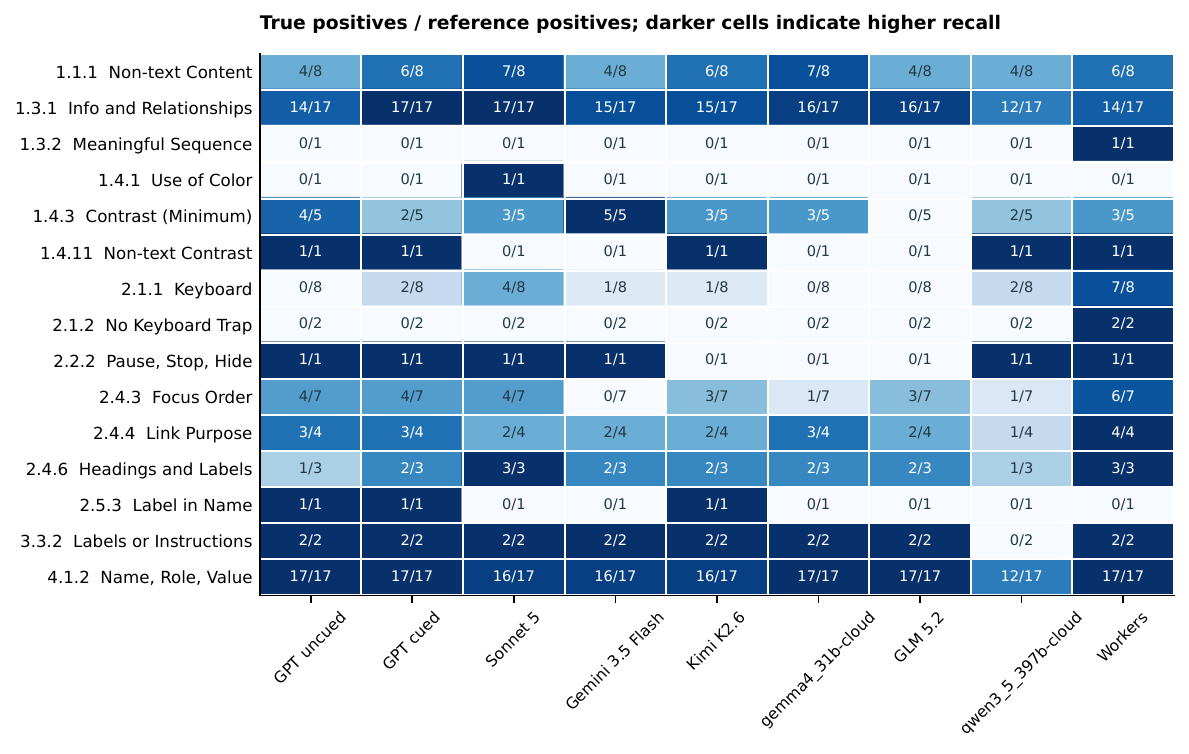}
\caption{Detection across all criteria with positive reference cases. Each cell reports true positives over reference positives; shading encodes recall. The worker column uses the reference run. Criteria with no positive examples are omitted because recall is undefined.}
\Description{A heatmap covers fifteen criteria, GPT uncued and cued batch conditions, six further batch-model configurations, and the reference worker run. All batch columns show zero of two for No Keyboard Trap, while Keyboard and Focus Order contain varying nonzero counts. Every cell prints its own positive denominator.}
\label{fig:critmodel}
\end{figure}

Figure~\ref{fig:critmodel} makes this variation visible without treating a result on one positive as equivalent evidence to a result on seventeen. Each cell contains the number of recovered reference cases and its denominator. The same shading scale is used throughout, while the counts make small samples explicit. The additional uncued GPT column separates the role of the batch cues from differences between the recorded model configurations.

On the same 250 rows, instrumented workers using Gemini 3.1 Pro, Claude Sonnet 5, and GPT-5.5 recover 60, 62, and 66 of 78 positives. A GPT-5.5 repeat recovers 69; the two passes agree on \repeatSame{} verdicts (\repeatAgreement{}\%). This two-pass comparison provides a limited repeatability check, with separate predictions preserving the changed rows for inspection.

\subsection{Costs attached to their own runs}
Table~\ref{tab:costs} pairs accuracy and cost for each complete archived workload. The uncued and cued batch VLM estimates are USD \costPure{} and USD \costCued{}, respectively, while the per-criterion estimate is USD \costPc{}. Batch processing shares page context across criteria, whereas the per-criterion procedure sends a separate assessment request for each row. These totals therefore describe different allocations of model calls as well as different detection outcomes. The per-criterion condition costs more while abstaining frequently; its higher conditional recall should be considered together with both this cost and the findings it leaves unresolved.

\begin{table}[t]
\centering
\small
\caption{Costs paired with accuracy from the same complete 250-row workload. Batch and coding conditions reuse page-level calls. Costs are USD estimates, not uniform billing measurements.}
\label{tab:costs}
\begin{tabular}{lrrrr}
\toprule
Condition & P & R & F1 & USD \\
\midrule
Batch VLM, uncued & 0.61 & 0.67 & 0.64 & 6.38 \\
Batch VLM, cued & 0.64 & 0.74 & 0.69 & 7.29 \\
Per-criterion VLM & 0.62 & 0.59 & 0.61 & 14.98 \\
Claude Code, naive & 0.64 & 0.60 & 0.62 & 58.60 \\
Claude Code, parity & 0.55 & 0.77 & 0.64 & 74.52 \\
Codex, naive & 0.71 & 0.32 & 0.44 & 32.76 \\
Codex, parity & 0.64 & 0.54 & 0.58 & 38.63 \\
Gemini 3.1 Pro & 0.57 & 0.77 & 0.66 & 36.65 \\
Claude Sonnet 5 & 0.55 & 0.79 & 0.65 & 49.71 \\
GPT-5.5 & 0.55 & 0.85 & 0.67 & 64.19 \\
GPT-5.5 repeat & 0.56 & 0.88 & 0.69 & 64.90 \\
\bottomrule
\end{tabular}
\end{table}

\begin{figure}[t]
\centering
\includegraphics[width=\linewidth]{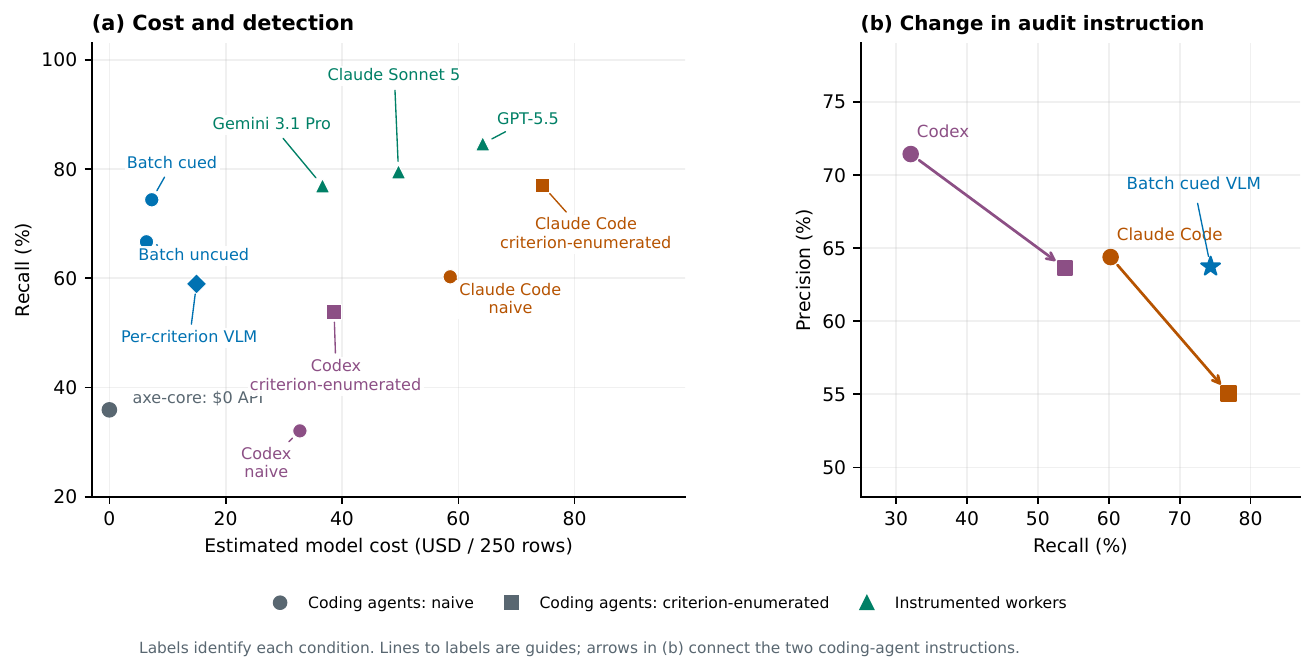}
\caption{Cost and precision--recall tradeoffs on 250 records. Worker triangles use the instrumented runs and their own usage records; the reference worker run is omitted from the cost panel because matching usage is unavailable. Coding-agent costs are API-equivalent estimates. The axe point denotes zero model API cost, not zero computational or human cost.}
\Description{The left scatterplot uses a linear cost axis, including zero, against recall. Labels and guide lines identify individual conditions; triangles denote instrumented workers and a diamond denotes the per-criterion VLM. The right plot connects each coding agent's naive circle to its criterion-enumerated square; both arrows move toward higher recall and lower precision.}
\label{fig:tiers}
\end{figure}

Figure~\ref{fig:tiers} shows the resource tradeoff directly. The left panel places each measured configuration at its own recall and estimated model cost. The right panel connects the two instructions within each coding-agent product, showing that more reference findings can accompany lower precision. The arrows identify the paired requests within each product.

The September 2026 worker estimates range from USD \costGemini{} for Gemini 3.1 Pro to USD \costGpt{} for the first GPT-5.5 pass, with their corresponding accuracy shown in the same rows. The reference worker result has no verified matching usage record, so no cost is assigned to it. Coding-agent figures range from USD \costCodexNaive{} to USD \costClaudeCodeParity{} across the recorded prompts and products. Those sessions use subscription authentication, and the reported amounts are API-equivalent usage estimates rather than evidence of cash charged for each audit. Comparing these resource estimates requires retaining their run identities and accounting methods.

Timing records require similar care. The first September 2026 GPT-5.5 worker run contains 19,509.9 seconds of summed within-row processing spans, approximately 78.0 seconds per timed row. This includes work between the logged first and last worker steps and is distinct from the sum of API response latencies. The archived timing procedure excludes spans over one hour to avoid counting long interruptions as active processing and records how many spans remain. Summing row durations also does not give the elapsed duration of a concurrent batch. We retain these quantities in the evidence records and avoid a single latency ranking that would merge page-level calls, row-level processing, and concurrent execution.

\subsection{Complementarity and sensitivity analyses}
\begin{figure}[t]
\centering
\includegraphics[width=\linewidth]{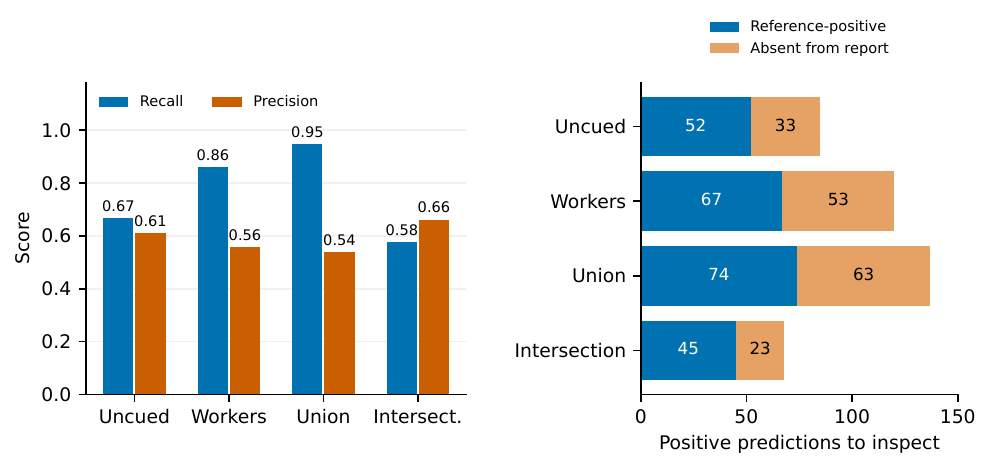}
\caption{Combining uncued VLM and reference-worker predictions. The union recovers additional reference positives but also adds positive predictions absent from the reports; requiring agreement reduces both. These are retrospective set operations, not a measured auditor workflow.}
\Description{Grouped bars show precision and recall for the uncued VLM, reference workers, union, and intersection. Stacked horizontal bars separate true-positive and reference-negative predictions that would require inspection. The union contains 74 true positives and 63 false positives; the intersection contains 45 and 23.}
\label{fig:complementarity}
\end{figure}

Combining historical worker and uncued VLM predictions illustrates a possible detection tradeoff. Figure~\ref{fig:complementarity} shows both the metric tradeoff and the number of positive predictions implied by each combination. Their union recovers \unionTP{} of \npos{} positives, with \unionFP{} false positives against the reference labels, yielding precision \unionP{} and recall \unionR{}. Their intersection returns \intersectionTP{} true positives and \intersectionFP{} false positives, yielding precision \intersectionP{} and recall \intersectionR{}. These retrospective set operations leave escalation, information sharing, and auditor adjudication for a subsequent workflow evaluation. The union shows remaining nonoverlap between the predictions, while the intersection shows the findings lost by requiring agreement.

Excluding development-exposed pages leaves \nheldrows{} rows and \nheldpos{} positives; workers recover 14 positives, the uncued VLM ten, and axe six (Table~\ref{tab:sensitivity}). The recall pattern persists, but the small retrospective subset cannot establish transfer to future deployments. Shared instructions revised on excluded pages may still affect retained pages. This sensitivity checks whether results concentrate on documented development pages, rather than providing a prospectively separated test set.

\begin{table}[t]
\centering
\small
\caption{Descriptive sensitivity analyses of the full benchmark. P/R entries use each row population consistently across conditions. Development exclusion removes whole pages; it does not establish an independent test set.}
\label{tab:sensitivity}
\begin{tabular}{lrrrrr}
\toprule
Population & Rows & Pos. & axe P/R & Uncued P/R & Workers P/R \\
\midrule
Primary & 250 & 78 & 0.90/0.36 & 0.61/0.67 & 0.56/0.86 \\
Development pages excluded & 55 & 15 & 1.00/0.40 & 0.62/0.67 & 0.58/0.93 \\
Axe-sourced rows excluded & 239 & 67 & 0.86/0.28 & 0.57/0.64 & 0.53/0.88 \\
Dynamic-only rows excluded & 240 & 68 & 0.89/0.35 & 0.59/0.69 & 0.53/0.88 \\
Deque reports & 103 & 33 & 1.00/0.39 & 0.65/0.61 & 0.59/0.82 \\
Accessiblu reports & 147 & 45 & 0.83/0.33 & 0.59/0.71 & 0.54/0.89 \\
\bottomrule
\end{tabular}
\end{table}

Vendor-specific results similarly show higher worker recall and lower precision than axe in both report subsets. Excluding rows whose provenance includes axe changes the composition of the positives and is reported with its own denominator. Excluding dynamic-only flags removes cases for which the archive records a mismatch between the required behavior and snapshot capabilities. These populations answer different questions, so none is selected as a replacement headline on the basis of better performance. Across leave-one-platform-out analyses, historical worker recall ranges from 0.84 to 0.90, compared with 0.64 to 0.70 for the uncued VLM and 0.33 to 0.37 for axe. These ranges describe sensitivity to platform composition; they are not confidence intervals or significance tests.

\section{Failure Analysis}
\label{sec:failure}
Archived diagnostic flags describe possible additional findings, unavailable saved-page behavior, worker defects, uncertain criterion assignments, and differences between reported and captured page states. The \texttt{bundled\_sc} flag indicates that a report description may concern a different criterion; \texttt{stale\_label} indicates that a reported feature appears different in the saved page. These annotations are not independently confirmed label corrections or an independent coding study. Their populations also differ: a dataset-wide observability flag is not automatically a worker false-positive category. Supplementary records preserve each flag alongside its prediction outcome.

Of the reference workers' 53 positive predictions on negative-labelled rows, 27 carry candidate-discovery flags. The authors jointly reviewed generated assessment reports, screenshots, and execution logs, without directly inspecting saved pages. These candidates lack independent adjudication; we neither amend primary labels nor report inter-rater reliability. A candidate could reflect an omitted finding, changed page, incorrect interpretation, or criterion mismatch. Resolving these alternatives requires individual evidence, rather than the detector's confidence or absence of a report entry.

We quantify the consequence of one explicit assumption: if all 27 flagged candidates were accepted as additional positives, historical worker precision would be \candidatePrecision{} and recall \candidateRecall{} on the revised label set. The precision denominator remains the worker's positive predictions, while the recall denominator grows from \npos{} to \candidatePos{} positive labels. We apply the same hypothetical label changes to all comparison conditions in the supplementary analysis. This calculation is a conditional sensitivity, not a second validated performance estimate or a guaranteed bound on true accuracy. The original reference labels remain unchanged in the primary analysis, and the candidate list is retained for future adjudication.

The remaining disagreements also require distinguishing evidence acquisition from interpretation. A tool may fail to reach a relevant state; a saved document may not contain it; or a worker may reach it and draw an unsupported conclusion. The OpenStax example illustrates another distinction: a correct page--criterion prediction need not identify the same issue as the source report. Aggregate confusion counts cannot establish how often each mechanism occurs. We therefore avoid assigning all residual false positives to ambiguous normative judgments or all false negatives to inadequate probing. The archived records motivate targeted review of those mechanisms, but do not support a complete independently validated taxonomy of error causes.

The widget probe's repeated-focus summary illustrates this distinction. A radio option can legitimately retain focus when selected; No Keyboard Trap instead concerns whether appropriate keyboard actions can move focus away~\cite{keyboardtrap,radiopattern}. A convincing trace must identify the component, attempted exit actions, and resulting focus behavior. We retain the archived criterion-level predictions, but focus retention alone cannot validate an issue-level violation: the evidence must demonstrate the criterion's failure condition.
\section{Discussion}
\label{sec:discussion}
\subsection{Matching assessment methods to evidence requirements}
The results suggest treating evidence collection as a design decision within accessibility auditing. The workers in the reference run recover more of the keyboard reference cases than the noninteractive VLM configurations, while some visual or semantic criteria favor a VLM condition. This pattern is consistent with prior work showing that different evaluation procedures expose different aspects of accessibility~\cite{vigo2013,bagel2023,gena11y2025}. It supports a criterion-oriented account of assessment, provided that the account distinguishes the evidence needed for a particular check from an absolute classification of an entire criterion. A single requirement may admit several failure modes, and an operation that exposes one of them may say little about the others.

For framework design, the practical consequence is to specify what an observation establishes. A DOM attribute can support a structural check; a screenshot can support interpretation of appearance; and a keyboard trace can document a transition between interface states. A worker that combines them should retain those distinctions in its report. This follows the emphasis on execution evidence in systems such as Groundhog and TaskAudit~\cite{groundhog2022,zhong2026taskaudit}. It also suggests a useful extension to the present implementation: record which evidential requirement was satisfied and which remained unresolved for each assessment. That extension could make partial coverage more legible, although its usefulness to auditors would need to be tested.

Abstentions can prevent unsupported verdicts, but the per-criterion VLM's high conditional recall excludes declined rows. Deployments must retain these as outstanding work. Human--AI interaction guidelines recommend communicating system capabilities~\cite{amershi2019}; here, a useful limitation statement would identify the missing observation and next inspection, rather than attach a generic confidence label.

\subsection{What report-derived benchmarks can establish}
Reusing professional audit reports connects evaluation to problems documented in an existing accessibility practice. It also requires preserving the scope of that evidence. WCAG-EM situates findings within an evaluation process that includes sampling and reporting~\cite{wcagem,wcagem2}; it does not imply that absence from a report is an exhaustive negative judgment. The LAA itself describes its evaluations as time-bounded assessments that identify some, but not all, barriers~\cite{laa2024}. Our candidate-discovery sensitivity illustrates how that difference can affect measured precision. The appropriate response is to expose the inference used to construct negative labels, retain the original labels, and identify proposed amendments as claims requiring adjudication. Accepting a model's disputed findings as new ground truth would weaken the independence of the evaluation it is supposed to improve.

The issue extends beyond whether an additional finding is correct. A saved page can differ from the service originally examined, and a criterion-level label can aggregate several distinct problems. These are differences in the object of assessment and its granularity, not simply annotation noise. Future datasets could preserve the audited page state, the implicated elements or interaction sequence, and explicit findings about criteria that were checked without a violation. Such records would make it possible to evaluate whether an issue exists, where it occurs, and whether the observations substantiate it separately. Mutation-based evaluation offers a complementary way to introduce known defects: Ma11y inserts WCAG-derived failures into rendered pages and checks whether tools detect them~\cite{ma11y2024}. Combining such tests with report-derived cases could distinguish missed seeded defects from uncertain labels in naturally occurring cases. Our manifest and row-level records provide part of that structure while keeping unavailable historical information explicitly unresolved.

Even a more exhaustive criterion benchmark would not replace studies of interface use. Power et al. show that guideline-based evaluation misses barriers encountered by blind users~\cite{power2012}. The LAA-derived corpus evaluates an automated assessment task; establishing improved accessibility of scholarly services additionally requires examining remediation, user outcomes, and how professional decisions connect findings to development work.

\subsection{Allocating automated effort within an audit}
The combinations of precision, recall, and resource use motivate a possible tiered workflow: run appropriate deterministic checks, apply model-based assessment where contextual interpretation is needed, and request targeted browser operations for unresolved cases. This proposal builds on complementary evaluation methods and on work addressing the scalability of professional auditing~\cite{vigo2013,gu2025scalable}. The ordering and routing policy require evaluation. Escalation should depend on the requirement and available evidence: an inexpensive stage may miss a case that requires further assessment.

The union and intersection analyses indicate why routing deserves separate evaluation. Taking every finding from either method increases recovery of reference positives but also produces more findings for review. Requiring agreement discards some positive reference cases that only one method finds. A deployed policy must account for these consequences alongside the cost of obtaining the second assessment. It must also consider correlation: workers can call deterministic tools, and cued VLMs receive heuristic information, so outputs are not independent votes. The next empirical question is how an explicit allocation rule performs under a stated budget on a separately selected population, with unresolved cases retained in the outcome measures.

A recent preprint by Oyelayo et al. reports that iterative LLM-based repair increased cost without improving remediation outcomes in its tested setting~\cite{oyelayo2026repair}. Although repair differs from detection, that finding reinforces the need to evaluate the benefit of additional calls rather than assume that iteration improves results. Resource estimates alone are insufficient for evaluating that workflow. Bi et al.'s account of accessibility in software practice situates tools within organizational constraints and development activities~\cite{bi2022}. An additional model call can be inexpensive relative to the effort required to reproduce an unclear finding or communicate it to a developer. Conversely, a more costly execution record may be useful if it makes a disputed state straightforward to inspect. Our study measures neither of those human costs. A subsequent evaluation should therefore include review time, adjudication accuracy, reproducibility of findings, and the work needed to act on them, rather than optimizing model expenditure in isolation.

\subsection{Inspectable evidence for human auditors}
The worker architecture produces records that could support professional review by connecting a criterion to operations and observations. A useful auditor-facing presentation would distinguish the observed element or state, the relevant requirement, and the model's interpretation. It should also preserve contrary evidence, failed operations, and the point at which execution stopped. AXNav's emphasis on replay provides a precedent for treating execution as material that people can inspect~\cite{axnav2024}. Our trace example demonstrates the availability of a sequence, while leaving open whether the current log format is sufficient for a professional to reproduce and assess the finding efficiently.

Appropriate reliance depends on how auditors assess the evidence presented. Bu\c{c}inca et al. show that explanations can be insufficient to prevent overreliance, and Bansal et al. investigate the conditions under which combined human--AI performance exceeds the component performances~\cite{bucinca2021,bansal2021}. For accessibility auditing, a fluent explanation may be especially persuasive when it uses the terminology of a normative criterion. An interface should therefore make the supporting observation inspectable rather than requiring the auditor to accept the explanation's account of it. The erroneous interpretation of radio-button focus retention is an example of the kind of distinction that such inspection must expose.

A human-review study could compare verdict-only reports, explanatory reports, and reports linked to execution evidence. Outcomes should include rejection of incorrect findings, reproduction of genuine issues, and review effort, accounting for accessibility expertise: traces useful to specialists may burden other reviewers. Such a study would test whether the records' inspectability translates into improvements in auditing practice.
\section{Limitations and Future Work}
The study is bounded by its archival population and reference-label construction. The 24 pages represent scholarly platforms rather than a representative sample of web applications, and only 15 criteria have positive cases. Inferred negatives, high-level reports, bundled findings, and changes between audited services and saved documents can each affect agreement. The candidate sensitivity does not replace independent adjudication. Moreover, page--criterion scoring measures whether a criterion is flagged, not whether every reported element, explanation, or proposed cause is correct. Future evaluation should include issue-level matching and direct validation of the recorded evidence, alongside populations with explicit negative judgments.

The comparison concerns the tested configurations on one saved-page corpus. The available records do not reconstruct every historical execution setting or missing usage measurement. Tool access, instructions, system prompts, and some underlying models also differ, preventing causal isolation of browser interaction or prompt granularity. Future comparisons should preserve identical captured inputs and executable configurations at run time, vary one component where a causal question requires it, and archive usage with the resulting predictions.

Development exposure and clustering further limit generalization. Diagnostic files touch most evaluated pages, and shared changes can affect untouched rows. The whole-page exclusion sensitivity is small and retrospective. A prospectively separated evaluation would provide stronger evidence about transfer to new pages and platforms. Repeated execution is also limited to two passes of one September 2026 worker configuration. The observed agreement does not estimate stability for all conditions or all future model versions. Likewise, logged prices and API-equivalent usage estimates describe the archived workloads rather than stable deployment costs; no human review or remediation cost was measured.

The framework operates saved documents, whose behavior can differ from production services requiring authentication, network state, or multi-page processes. Extending the environment to such processes would introduce additional questions about state restoration, comparable inputs, and what constitutes sufficient evidence for an assessment. Separately, a study involving auditors and people with relevant access requirements is needed to evaluate the proposed workflow's practical consequences. These extensions concern different outcomes: more complete execution coverage, better-validated findings, and improved accessibility in use. Progress on one should not be assumed to establish the others.

\section{Conclusion}
We presented a framework that combines shared browser tools with criterion-specific instructions for accessibility assessment. On the report-derived benchmark, reference workers recover more positive labels than axe-core and an uncued batch VLM, with lower precision; detection patterns also vary across criteria. These findings motivate matching assessment procedures to the evidence a requirement needs. The framework preserves operations and observations, making the basis of a judgment available for inspection. Although benefits to professional auditing remain to be evaluated, criterion-specific evidence collection and inspectable execution records provide concrete directions for developing accessibility assessment systems.

\bibliographystyle{plainnat}

\begin{thebibliography}{31}
\providecommand{\natexlab}[1]{#1}
\providecommand{\url}[1]{\texttt{#1}}
\expandafter\ifx\csname urlstyle\endcsname\relax
  \providecommand{\doi}[1]{doi: #1}\else
  \providecommand{\doi}{doi: \begingroup \urlstyle{rm}\Url}\fi

\bibitem[Amershi et~al.(2019)Amershi, Weld, Vorvoreanu, Fourney, Nushi,
  Collisson, Suh, Iqbal, Bennett, Inkpen, Teevan, Kikin-Gil, and
  Horvitz]{amershi2019}
Saleema Amershi, Dan Weld, Mihaela Vorvoreanu, Adam Fourney, Besmira Nushi,
  Penny Collisson, Jina Suh, Shamsi Iqbal, Paul~N. Bennett, Kori Inkpen, Jaime
  Teevan, Ruth Kikin-Gil, and Eric Horvitz.
\newblock {Guidelines for Human-AI Interaction}.
\newblock In \emph{Proceedings of the 2019 CHI Conference on Human Factors in
  Computing Systems}, pages 1--13, New York, NY, USA, 2019. ACM.
\newblock \doi{10.1145/3290605.3300233}.

\bibitem[Bansal et~al.(2021)Bansal, Wu, Zhou, Fok, Nushi, Kamar, Ribeiro, and
  Weld]{bansal2021}
Gagan Bansal, Tongshuang Wu, Joyce Zhou, Raymond Fok, Besmira Nushi, Ece Kamar,
  Marco~Tulio Ribeiro, and Daniel Weld.
\newblock {Does the Whole Exceed its Parts? The Effect of AI Explanations on
  Complementary Team Performance}.
\newblock In \emph{Proceedings of the 2021 CHI Conference on Human Factors in
  Computing Systems}, pages 1--16, New York, NY, USA, 2021. ACM.
\newblock \doi{10.1145/3411764.3445717}.

\bibitem[Bassi et~al.(2025)Bassi, Delnevo, Franco, Gaggi, Gatto, Mirri, and
  Olaiya]{bassi2025}
Barry Bassi, Giovanni Delnevo, Mirko Franco, Ombretta Gaggi, Salvatore Gatto,
  Silvia Mirri, and Kelvin Olaiya.
\newblock {Supporting Accessibility Auditing and HTML Validation using Large
  Language Models}.
\newblock In \emph{Proceedings of the 40th ACM/SIGAPP Symposium on Applied
  Computing}, pages 27--31. Association for Computing Machinery, 2025.
\newblock \doi{10.1145/3672608.3707912}.

\bibitem[Bi et~al.(2022)Bi, Xia, Lo, Grundy, Zimmermann, and Ford]{bi2022}
Tingting Bi, Xin Xia, David Lo, John Grundy, Thomas Zimmermann, and Denae Ford.
\newblock {Accessibility in Software Practice: A Practitioner’s Perspective}.
\newblock \emph{ACM Transactions on Software Engineering and Methodology},
  31\penalty0 (4):\penalty0 1--26, 2022.
\newblock \doi{10.1145/3503508}.

\bibitem[Buçinca et~al.(2021)Buçinca, Malaya, and Gajos]{bucinca2021}
Zana Buçinca, Maja~Barbara Malaya, and Krzysztof~Z. Gajos.
\newblock {To Trust or to Think: Cognitive Forcing Functions Can Reduce
  Overreliance on AI in AI-assisted Decision-making}.
\newblock \emph{Proceedings of the ACM on Human-Computer Interaction},
  5\penalty0 (CSCW1):\penalty0 1--21, 2021.
\newblock \doi{10.1145/3449287}.

\bibitem[Chiou et~al.(2021)Chiou, Alotaibi, and Halfond]{keyboard2021}
Paul~T. Chiou, Ali~S. Alotaibi, and William G.~J. Halfond.
\newblock {Detecting and localizing keyboard accessibility failures in web
  applications}.
\newblock In \emph{Proceedings of the 29th ACM Joint Meeting on European
  Software Engineering Conference and Symposium on the Foundations of Software
  Engineering}, pages 855--867, New York, NY, USA, 2021. ACM.
\newblock \doi{10.1145/3468264.3468581}.

\bibitem[Chiou et~al.(2023{\natexlab{a}})Chiou, Alotaibi, and
  Halfond]{dialog2023}
Paul~T. Chiou, Ali~S. Alotaibi, and William G.~J. Halfond.
\newblock {Detecting Dialog-Related Keyboard Navigation Failures in Web
  Applications}.
\newblock In \emph{2023 IEEE/ACM 45th International Conference on Software
  Engineering (ICSE)}, pages 1368--1380, Melbourne, Australia,
  2023{\natexlab{a}}. IEEE.
\newblock \doi{10.1109/icse48619.2023.00120}.

\bibitem[Chiou et~al.(2023{\natexlab{b}})Chiou, Alotaibi, and
  Halfond]{bagel2023}
Paul~T. Chiou, Ali~S. Alotaibi, and William~G.J. Halfond.
\newblock {BAGEL: An Approach to Automatically Detect Navigation-Based Web
  Accessibility Barriers for Keyboard Users}.
\newblock In \emph{Proceedings of the 2023 CHI Conference on Human Factors in
  Computing Systems}, pages 1--17, New York, NY, USA, 2023{\natexlab{b}}. ACM.
\newblock \doi{10.1145/3544548.3580749}.

\bibitem[{Deque Systems}(2026)]{axecore}
{Deque Systems}.
\newblock {axe-core: Accessibility Engine for Automated Web UI Testing}, 2026.
\newblock URL \url{https://github.com/dequelabs/axe-core}.
\newblock Accessed September 8, 2026; software reference, not a claim about the
  historical run version.

\bibitem[Fathallah et~al.(2025)Fathallah, Hern\'{a}ndez, and
  Staab]{fathallah2025accessguru}
Nadeen Fathallah, Daniel Hern\'{a}ndez, and Steffen Staab.
\newblock {AccessGuru: Leveraging LLMs to Detect and Correct Web Accessibility
  Violations in HTML Code}.
\newblock arXiv:2507.19549, 2025.
\newblock URL \url{https://arxiv.org/abs/2507.19549}.

\bibitem[Gu et~al.(2026)Gu, Wang, Lai, Gao, Zhou, and Bu]{gu2025scalable}
Ming Gu, Ziwei Wang, Sicen Lai, Zirui Gao, Sheng Zhou, and Jiajun Bu.
\newblock {Towards Scalable Web Accessibility Audit with MLLMs as Copilots}.
\newblock \emph{Proceedings of the AAAI Conference on Artificial Intelligence},
  40\penalty0 (45):\penalty0 38515--38523, 2026.
\newblock \doi{10.1609/aaai.v40i45.41193}.

\bibitem[He et~al.(2025)He, Huq, and Malek]{gena11y2025}
Ziyao He, Syed~Fatiul Huq, and Sam Malek.
\newblock {Enhancing Web Accessibility: Automated Detection of Issues with
  Generative AI}.
\newblock \emph{Proceedings of the ACM on Software Engineering}, 2\penalty0
  (FSE):\penalty0 2264--2287, 2025.
\newblock \doi{10.1145/3729371}.

\bibitem[{Library Accessibility Alliance}(2026)]{laa2024}
{Library Accessibility Alliance}.
\newblock {Library Accessibility Alliance: E-resource Accessibility
  Evaluations}, 2026.
\newblock URL \url{https://www.libraryaccessibility.org/testing}.
\newblock Accessed September 8, 2026.

\bibitem[Lormeau(2026)]{singlefile}
Gildas Lormeau.
\newblock {SingleFile: Save a Complete Web Page as a Single HTML File}, 2026.
\newblock URL \url{https://github.com/gildas-lormeau/SingleFile}.
\newblock Accessed September 8, 2026; software documentation.

\bibitem[Oyelayo et~al.(2026)Oyelayo, Abushaqra, Asadi, Dey, and
  Costa]{oyelayo2026repair}
Oluwatoyosi Oyelayo, Ghada Abushaqra, Parham Asadi, Durjoy Dey, and Diego~Elias
  Costa.
\newblock {LLM Based Web Accessibility Repair: An Empirical Study of Detection,
  Remediation, and Cost}.
\newblock arXiv preprint arXiv:2605.27716, 2026.
\newblock URL \url{https://arxiv.org/abs/2605.27716}.

\bibitem[Power et~al.(2012)Power, Freire, Petrie, and Swallow]{power2012}
Christopher Power, André Freire, Helen Petrie, and David Swallow.
\newblock {Guidelines are only half of the story: accessibility problems
  encountered by blind users on the web}.
\newblock In \emph{Proceedings of the SIGCHI Conference on Human Factors in
  Computing Systems}, pages 433--442, New York, NY, USA, 2012. ACM.
\newblock \doi{10.1145/2207676.2207736}.

\bibitem[Salehnamadi et~al.(2022)Salehnamadi, Mehralian, and
  Malek]{groundhog2022}
Navid Salehnamadi, Forough Mehralian, and Sam Malek.
\newblock {Groundhog: An Automated Accessibility Crawler for Mobile Apps}.
\newblock In \emph{Proceedings of the 37th IEEE/ACM International Conference on
  Automated Software Engineering}, pages 1--12, New York, NY, USA, 2022. ACM.
\newblock \doi{10.1145/3551349.3556905}.

\bibitem[Taeb et~al.(2024)Taeb, Swearngin, Schoop, Cheng, Jiang, and
  Nichols]{axnav2024}
Maryam Taeb, Amanda Swearngin, Eldon Schoop, Ruijia Cheng, Yue Jiang, and
  Jeffrey Nichols.
\newblock {AXNav: Replaying Accessibility Tests from Natural Language}.
\newblock In \emph{Proceedings of the CHI Conference on Human Factors in
  Computing Systems}, pages 1--16, New York, NY, USA, 2024. ACM.
\newblock \doi{10.1145/3613904.3642777}.

\bibitem[Tafreshipour et~al.(2024)Tafreshipour, Deshpande, Mehralian, Ahmed,
  and Malek]{ma11y2024}
Mahan Tafreshipour, Anmol Deshpande, Forough Mehralian, Iftekhar Ahmed, and Sam
  Malek.
\newblock {Ma11y: A Mutation Framework for Web Accessibility Testing}.
\newblock In \emph{Proceedings of the 33rd ACM SIGSOFT International Symposium
  on Software Testing and Analysis}, pages 100--111. Association for Computing
  Machinery, 2024.
\newblock \doi{10.1145/3650212.3652113}.

\bibitem[Vigo et~al.(2013)Vigo, Brown, and Conway]{vigo2013}
Markel Vigo, Justin Brown, and Vivienne Conway.
\newblock {Benchmarking web accessibility evaluation tools: measuring the harm
  of sole reliance on automated tests}.
\newblock In \emph{Proceedings of the 10th International Cross-Disciplinary
  Conference on Web Accessibility}, pages 1--10, New York, NY, USA, 2013. ACM.
\newblock \doi{10.1145/2461121.2461124}.

\bibitem[{World Wide Web Consortium}(2014)]{wcagem}
{World Wide Web Consortium}.
\newblock {Website Accessibility Conformance Evaluation Methodology (WCAG-EM)
  1.0}.
\newblock W3C Working Group Note, 2014.
\newblock URL \url{https://www.w3.org/TR/2014/NOTE-WCAG-EM-20140710/}.

\bibitem[{World Wide Web Consortium}(2018)]{w3cwai}
{World Wide Web Consortium}.
\newblock {Web Content Accessibility Guidelines (WCAG) 2.1}.
\newblock W3C Recommendation, 2018.
\newblock URL \url{https://www.w3.org/TR/2018/REC-WCAG21-20180605/}.

\bibitem[{World Wide Web Consortium}(2023)]{wcag22}
{World Wide Web Consortium}.
\newblock {Web Content Accessibility Guidelines (WCAG) 2.2}.
\newblock W3C Recommendation, 2023.
\newblock URL \url{https://www.w3.org/TR/2023/REC-WCAG22-20231005/}.

\bibitem[{World Wide Web Consortium}(2026{\natexlab{a}})]{keyboardtrap}
{World Wide Web Consortium}.
\newblock {Understanding Success Criterion 2.1.2: No Keyboard Trap},
  2026{\natexlab{a}}.
\newblock URL
  \url{https://www.w3.org/WAI/WCAG21/Understanding/no-keyboard-trap.html}.
\newblock Accessed September 8, 2026.

\bibitem[{World Wide Web Consortium}(2026{\natexlab{b}})]{nontext}
{World Wide Web Consortium}.
\newblock {Understanding Success Criterion 1.1.1: Non-text Content},
  2026{\natexlab{b}}.
\newblock URL
  \url{https://www.w3.org/WAI/WCAG21/Understanding/non-text-content.html}.
\newblock Accessed September 8, 2026.

\bibitem[{World Wide Web Consortium}(2026{\natexlab{c}})]{radiopattern}
{World Wide Web Consortium}.
\newblock {ARIA Authoring Practices Guide: Radio Group Pattern},
  2026{\natexlab{c}}.
\newblock URL \url{https://www.w3.org/WAI/ARIA/apg/patterns/radio/}.
\newblock Accessed September 8, 2026.

\bibitem[{World Wide Web Consortium}(2026{\natexlab{d}})]{wcagem2}
{World Wide Web Consortium}.
\newblock {WCAG Evaluation Methodology (WCAG-EM) 2.0}.
\newblock W3C Group Note, July 23, 2026{\natexlab{d}}.
\newblock URL \url{https://www.w3.org/TR/2026/NOTE-wcag-em-2-20260723/}.

\bibitem[Yao et~al.(2023)Yao, Zhao, Yu, Du, Shafran, Narasimhan, and
  Cao]{react2023}
Shunyu Yao, Jeffrey Zhao, Dian Yu, Nan Du, Izhak Shafran, Karthik Narasimhan,
  and Yuan Cao.
\newblock {ReAct: Synergizing Reasoning and Acting in Language Models}.
\newblock International Conference on Learning Representations, 2023.
\newblock URL \url{https://arxiv.org/abs/2210.03629v3}.

\bibitem[Zheng et~al.(2026)Zheng, Lee, Benes, and Yeh]{li2026webaccessvl}
Amber~Yijia Zheng, Jae~Joong Lee, Bedrich Benes, and Raymond~A. Yeh.
\newblock {WebAccessVL: Violation-Aware VLM for Web Accessibility}.
\newblock arXiv:2602.03850, version 3, 2026.
\newblock URL \url{https://arxiv.org/abs/2602.03850v3}.

\bibitem[Zhong et~al.(2025)Zhong, Chen, Chen, Fogarty, and
  Wobbrock]{screenaudit2025}
Mingyuan Zhong, Ruolin Chen, Xia Chen, James Fogarty, and Jacob~O. Wobbrock.
\newblock {ScreenAudit: Detecting Screen Reader Accessibility Errors in Mobile
  Apps Using Large Language Models}.
\newblock In \emph{Proceedings of the 2025 CHI Conference on Human Factors in
  Computing Systems}, pages 1--19, New York, NY, USA, 2025. ACM.
\newblock \doi{10.1145/3706598.3713797}.

\bibitem[Zhong et~al.(2026)Zhong, Chen, Kyi, Li, Fogarty, and
  Wobbrock]{zhong2026taskaudit}
Mingyuan Zhong, Xia Chen, Davin~Win Kyi, Chen Li, James Fogarty, and Jacob~O.
  Wobbrock.
\newblock {TaskAudit: Detecting Functiona11ity Errors in Mobile Apps via
  Agentic Task Execution}.
\newblock In \emph{Proceedings of the 2026 CHI Conference on Human Factors in
  Computing Systems}, pages 1--20, New York, NY, USA, 2026. ACM.
\newblock \doi{10.1145/3772318.3791415}.

\end{thebibliography}

\end{document}